\documentclass[utf8]{frontiersSCNSmod} 

\usepackage{url,hyperref,lineno,microtype,subcaption}
\usepackage[onehalfspacing]{setspace}
\usepackage{bm}

\def\keyFont{\fontsize{8}{11}\helveticabold }
\def\firstAuthorLast{Juan Ruben Gomez-Solano} 
\def\Authors{Juan Ruben Gomez-Solano\,$^{1,*}$}
\def\Address{$^{1}$Departamento de Sistemas Complejos, Instituto de F\'isica, Universidad Nacional Aut\'onoma de M\'exico, Cd. de M\'exico, C.P. 04510, M\'exico}
\def\corrAuthor{Corresponding Author}

\def\corrEmail{r\_gomez@fisica.unam.mx}

\begin{document}
\onecolumn
\firstpage{1}

\title[Colloidal heat engines in viscoelastic baths]{Work extraction and performance of colloidal heat engines in viscoelastic baths} 

\author[\firstAuthorLast ]{\Authors} 
\address{} 
\correspondance{} 

\extraAuth{}

\maketitle

\begin{abstract}

\section{}
A colloidal particle embedded in a fluid can be used as a microscopic heat engine by means of a sequence of cyclic transformations imposed by an optical trap. We investigate a model for the operation of such kind of Brownian engines when the surrounding medium is viscoelastic, which endows the particle dynamics with memory friction. We analyze the effect of the relaxation time of the fluid on the performance of the colloidal engine under finite-time Stirling cycles. We find that, due to the frequency-dependence of the friction in viscoelastic fluids, the mean power delivered by the engine and its efficiency can be highly enhanced as compared to those in a viscous environment with the same zero-shear viscosity. In addition, with increasing fluid relaxation time the interval of cycle times at which positive power output can be delivered by the engine broadens. Our results reveal the importance of the transient behavior of the friction experienced by a Brownian heat engine in a complex fluid, which cannot be neglected when driven by thermodynamic cycles of finite duration.

\tiny
 \keyFont{ \section{Keywords:} stochastic thermodynamics, memory, viscoelasticity, heat engine, fluctuations, thermodynamic cycles, nonequilibrium process , memory effects} 
\end{abstract}

\section{Introduction}


Historically, the study of heat engines has played a fundamental role in the general understanding of energy exchanges in macroscopic systems. For instance, the conception of the well-known Carnot cycle almost two centuries ago was motivated by the design of efficient engines capable of performing mechanical work by extracting energy from a hot reservoir and transfering heat to a cold reservoir, which finally led to the formulation of the second law of thermodynamics. Carnot theorem imposes a universal bound for the maximum efficiency that can be ideally achieved by any heat engine working reversibly in the quasi-static limit. Since then, further theoretical results on the efficiency of irreversible heat engines under finite-time thermodynamic cycles with non-zero power output have been obtained~\citep{Novikov1958,Curzon1975,Leff1987,VanDenBroeck2005,Izumida2008,Esposito2009}, which turn out to be important for practical applications.

In more recent years, advances in miniaturization technologies have allowed researchers in both basic and applied science to conceive the design of micron- and submicron-sized machines with the ability to perform specific tasks in the mesoscopic realm, e.g. controlled cargo transport through microchannels and nanopores, in situ cell manipulation, assembly of functional microstructures, micropumping, microflow rectification, micromixing of fluids, and bio-inspired artificial locomotion~\citep{Ozin2005,Hanggi2009,Kim2016}. This has triggered an increasing interest in investigating the energetics and performance of mesoscopic heat engines, which, similar to their macroscopic counterparts, must be able to convert in an efficient manner the energy absorbed from their environment into useful work~\citep{Martinez2017,Pietzonka2019}. An important issue that arises in the theoretical description and implementation of such  devices is that  they must operate under highly non-equilibrium conditions with pronounced thermal fluctuations, which poses important conceptual and practical challenges~\cite{Ciliberto2013}. A significant progress in the theoretical analysis of mesoscopic heat engines has been made in the last two decades with the advent of stochastic thermodynamics, which extends concepts of classical thermodynamics such as heat, work and entropy production to the level of single stochastic trajectories for both equilibrium and driven systems~\citep{Sekimoto1998,Seifert2012,Speck2016,Ciliberto2017}. Within this theoretical framework, it is possible to carry out a comprehensive analysis of the performance of stochastic heat engines based on Brownian particles subject to periodically time-dependent potentials and temperatures~\citep{Schmiedl2007,Rana2014,Holubec2014,Tu2014,Bauer2016}. Along the same lines, optical micromanipulation techniques have facilitated during the last decade the experimental realization of simple colloidal heat engines, which are composed of a single colloidal particle as a working substance, embedded in water as a heat reservoir, undergoing thermodynamic cycles controlled by a harmonic optical potential~\citep{Blickle2012,QuintoSu2014,Martinez2016,Argun2017,Albay2021}. In such colloidal systems, expansions and compresions during Stirling- and Carnot-like cycles are achieved by decreasing and increasing the trap stiffness, respectively, while a hot reservoir is realized either by an actual increase of the local temperature of the around the particle or by addition of synthetic noise of non-thermal origin. These experiments have paved the way for the investigation of stochastic models of colloidal heat engines in more intricate and realistic situations, such as passive Brownian engines operating in contact with active baths~\citep{Zakine2017,Saha2018,Chaki2018,Chaki2019,Saha2019,Holubec2020}, Brownian engines with a self-propelled particle as working substance in contact with a viscous fluid~\cite{Ekeh2020,Kumari2020,Szamel2020,Holubec2020} or in a suspension of passive Brownian particles~\citep{Martin2018} as a heat bath, as well as the realization of a colloidal Stirling engine in bacterial baths with tunable activity \citep{Krishnamurthy2016}. 

It must be pointed out that, in most of the situations envisaged for biological and technological applications, the fluid environment of a colloidal heat engine is not perfectly Newtonian with a contant viscosity, but possesses a complex viscoelastic microstructure because of the presence of macromolecules, e.g., biomolecular chains, polymers and wormlike micelles, or colloids suspended in a solvent, thus exhibiting time-dependent flow properties~\cite{Larson1999}. Therefore, the motion of a colloidal particle in such materials lacks a clear-cut separation from timescales of the surroundings, which results in memory effects with large relaxation times. All these features give rise to a wealth of intriguing transient effects that markedly manifest themselves when time-dependent driving forces are exerted on an embedded particle~\citep{Wilson2011,Demery2014,GomezSolano2014,GomezSolano2015,Berner2018,Mohanty2020}, and are absent in the case of purely viscous fluids. Although all these conditions are met by a colloidal heat engine operating in a complex fluid, to the best of our knowledge they have never been examined in the context of stochastic thermodynamic cycles. Therefore, it is of paramount importance to assess the role of viscoelasticity in the performance of this kind of engines, since the resulting frequency-dependent friction experienced by a colloidal particle can significantly impact the rate at which energy is dissipated into a viscoelastic bath~\citep{Toyabe2008,Vishen2020,DiTerlizzi2020,Di_Terlizzi2020}.  

Here, we investigate a model based on the generalized Langevin equation for the operation of a stochastic Stirling engine composed of a Brownian particle embedded in a viscoelastic fluid bath, which includes a memory kernel and colored noise to account for retarded friction effects and thermal fluctuations of the medium on the particle motion. By numerically solving the correspoding non-Markovian equation of motion, we analyze the effect of the characteristic relaxation time of the fluid on the performance of the engine under finite-time Stirling cycles, and compare our results with those found in the case of Brownian particle in a Markovian bath. We uncover a significant increase in the power output and the efficiency of the engine operating in a viscoelastic environment with respect to the corresponding values in a viscous bath at a given cycle time. Moreover, with increasing relaxation time of the fluid, the convergence to the quasi-static Stirling efficiency is shifted to monotonically decreasing values of the cycle period, thereby expanding the interval at which the engine is able to efficiently deliver positive power.

\section{Model}

We consider a stochastic heat engine consisting of a Brownian particle embedded in a viscoelastic fluid as a heat bath, whose motion is confined by a harmonic potential. Both the curvature of the confining potential and the temperature of the system can be varied in time according to a well-specified periodic protocol that mimics a macroscopic thermodynamic cycle. Therefore, a stochastic model of the particle dynamics that allows for temporal variations of the temperature is needed. Based on Zwanzig's pioneering work~\citep{Zwanzig1973}, Brey et al.,~\citep{brey1990} and Romero-Salazar et al.~\citep{romerosalazar1995} derived the simplest equations of motion of a Brownian particle coupled to a heat bath with temperature changing in time. Their approach incorporates linear dissipative terms in the equations of motion of the surrounding bath particles, which account for continuous cooling or heating of the system controlled by some external mechanism in such a way that the bath particles are always in a canonical equilibrium at a well-behaved temperature dependent on time. In particular, in one dimension the generalized Langevin equation for the position $x(t)$ at time $t >0$ of the Brownian particle subject to a potential $U(x(t),t)$, reads~\citep{brey1990,romerosalazar1995} 
\begin{equation}\label{eq:GLEnonstatT0}
  m\frac{d^2 x(t)}{dt^2} = - \int_{0}^t ds \, K(t-s)\frac{d}{ds}\left[ \sqrt{\frac{T(t)}{T(s)}}x(s)\right] - \frac{dU(x(t),t)}{dx} +\zeta(t),
\end{equation}
where $m$ is the mass of the particle,  $T(s)$ is the temperature of the system at time $0 \le s \le t$, and $K(t-s)$ is a memory kernel that weights the effect of the previous history of the particle motion at time $s$ on its current drag force at time $t$ due to the temporal correlations induced by the surrounding medium. In addition, in Equation~(\ref{eq:GLEnonstatT0}), $\zeta(t)$ is a Gaussian stochastic force which accounts for thermal fluctuations in the system and satisfies 
\begin{eqnarray}\label{eq:colorednoise}
    \langle \zeta (t)\rangle & = & 0,\nonumber\\
    \langle \zeta(t) \zeta(s) \rangle & = & k_B \sqrt{T(t) T(s)} K(|t-s|).
\end{eqnarray}
Extensions of Equation~(\ref{eq:GLEnonstatT0}) to the three dimensional case, ${\bm{r}} = (x,y,z)$, which are relevant in many experimental situations using optical trapping techniques~\citep{gieseler2020}, are possible by a proper choice of the potential $U({\bm{r}},t)$ and a tensorial form of the memory kernel for particles of arbitrary shape~\citep{Squires2010}. Here, for the sake of simplicity we focus on the dynamics of a single coordinate of a spherical particle of radius $a$, which is confined by a harmonic potencial $U(x,t) = \frac{1}{2}\kappa(t) x(t)^2$, where $\kappa(t)$ is the stiffness at time $t$ of the corresponding restoring force. Moreover, we assume that the fluid bath is incompressible and the time-dependent variation of $\kappa(t)$ and $T(t)$ are such that its rheological properties remain in the linear viscoelastic regime, which is completely characterized by the stress relaxation modulus $G(t)$, or equivalently, by the complex dynamic shear modulus at frequency $\omega >0$, $G^*(\omega) = i\omega \eta^*(\omega)$, where $i = \sqrt{-1}$ and $ \eta^*(\omega)$ is the complex viscosity given by the Fourier transform of $G(t)$, {i.e.}, $\eta^*(\omega) = \int_{-\infty}^{\infty} dt\, e^{-\mathrm{i}\omega t} G(t) $~\citep{Bird1987}. In general, $G(t)$ is a function that decays to zero over a finite time-scale whose value  is many orders of magnitude greater than those of simple viscous fluids~\citep{Larson1999}. For $a$ larger than the characteristic length-scales of the fluid microstructure, the Fourier transform of the memory kernel, $\hat{K}(\omega) = \int_{-\infty}^{\infty} dt\, e^{-\mathrm{i}\omega t} K(t) $, is related to $\eta^*(\omega)$ by the generalized Stokes relation~\citep{Felderhof2009,Indei2012}
\begin{equation}\label{eq:GSR}
	\hat{K}(\omega) = 6\pi a \eta^*(\omega) \left[ 1 + a\sqrt{\frac{i\rho \omega}{\eta^*(\omega)}} \right],
\end{equation} 
with $\rho$ the density of the fluid. Furthermore, when $a$ is much smaller than the so-called viscoelastic penetration depth, $\sqrt{\frac{|\eta^*(\omega)|}{\rho \omega}}$, as typically occurs for micron-sized particles suspended in most viscoelastic fluids, inertial flow effects are negligible~\citep{Xu2007}. In such a case, Equation~(\ref{eq:GSR}) can be approximated to  $\hat{K}(\omega) = 6\pi a \eta^*(\omega)$~\citep{Cordoba2012}, which yields the simple relation  $K(t) = 6\pi a G(t)$ by Fourier inversion. This leads to the following Langevin equation for the position of the Brownian heat engine in the overdamped limit
\begin{equation}\label{eq:GLEnonstatT}
  6\pi a \int_{0}^t ds \, G(t-s)\frac{d}{ds}\left[ \sqrt{\frac{T(t)}{T(s)}}x(s)\right] =  - \kappa(t)x(t) +\zeta(t).
\end{equation}
In the following, we focus on a fluid relaxation modulus consisting of a Dirac delta function plus an exponential decay 
\begin{equation}\label{eq:relaxmodlexp}
    G(t) = 2\eta_{\infty} \delta(t) + \frac{\eta_{0}- \eta_{\infty}}{\tau_0} \exp \left( - \frac{t}{\tau_0} \right), t \ge 0,
\end{equation}
which models the rheological response of several viscoelastic fluids, such as wormlike micelles~\cite{Fischer1997,Ezrahi2006,GomezSolano2015}, some polymer solutions~\citep{Paul2019,Paul2021}, and to a great extent, the linear viscoelasticity over certain time intervals of intracellular fluids~\citep{Wilhelm2003,Vaippully2020}, block copolymers~\citep{Raspaud1996}, and $\lambda$-phage DNA~\cite{Zhu2008,GomezSolano2015}, where $\tau_0$ is the relaxation time of their elastic microstructure, whereas $\eta_0$ and $\eta_{\infty}$ represent the zero-shear viscosity and  the background solvent viscosity, respectively. Therefore, the corresponding friction memory kernel is
\begin{equation}\label{eq:kernelexp}
    K(t) = 2\gamma_{\infty} \delta(t) + \frac{\gamma_{0}- \gamma_{\infty}}{\tau_0} \exp \left( - \frac{t}{\tau_0} \right), t \ge 0,
\end{equation}
where the complex conjugate of its Fourier transform, $\hat{K}^*(\omega) = 6\pi a \eta(\omega) $, represents a frequency-dependent friction 
\begin{equation}\label{eq:Fourierkernelexp}
   \hat{K}^*(\omega) = \frac{\gamma_0 + \gamma_{\infty}\omega^2 \tau_0^2}{1 + \omega^2 \tau_0^2} + {i} \frac{(\gamma_0 - \gamma_{\infty})\omega \tau_0}{1 + \omega^2 \tau_0^2}.
\end{equation}
In Equations (\ref{eq:kernelexp}) and (\ref{eq:Fourierkernelexp}), $\gamma_{\infty} = 6\pi r \eta_{\infty}$ and $\gamma_0 = 6\pi r \eta_0 \ge \gamma_{\infty}$ are friction coefficients characterizing dissipation at short and long timescales, respectively, whereas elastic effects are quantified by $(\gamma_0 - \gamma_{\infty})\tau_0$. 
Hence, in this case Equation~(\ref{eq:GLEnonstatT}) takes the form
\begin{equation}\label{eq:GLEnonstatTexp}
 \gamma_{\infty}\frac{dx(t)}{dx} + \frac{\gamma_0 - \gamma_{\infty}}{\tau_0} \int_{0}^t ds \, \exp \left(-\frac{t-s}{\tau_0} \right)\frac{d}{ds}\left[ \sqrt{\frac{T(t)}{T(s)}}x(s)\right] = -\kappa(t) x(t) + \zeta(t).
\end{equation}
It is noteworthy that, at constant temperature $T$ and in absence of a trapping potential, the mean square displacement of a particle whose motion is described by Equation~(\ref{eq:GLEnonstatTexp}), is
\begin{equation}\label{eq:msdfree}
	\langle  \Delta x(t)^2 \rangle = \frac{2 k_B T}{\gamma_0} \left\{ t + \left(1-\frac{\gamma_{\infty}}{\gamma_0} \right)\tau_0 \left[ 1 - \exp\left(-\frac{\gamma_0}{\gamma_{\infty}\tau_0}t \right)\right] \right\},
\end{equation}
which implies that in the long-time limit, $t \gg \gamma_{\infty} \tau_0 / \gamma_0$, it would perform free diffusion like in a Newtonian fluid with constant viscosity $\eta_0$~\citep{Bellour2002,Grimm2011,Narinder2019}, i.e., $\langle  \Delta x(t)^2 \rangle \approx \frac{2k_B T}{\gamma_0}t$. This provides a clear criterion for a direct comparison of the performance of a Brownian engine in a viscoelastic fluid bath with that in a viscous medium of the same zero-shear viscosity, i.e., $\eta = \eta_0$, under identical time-dependent variations of $\kappa(t)$ and $T(t)$. Furthermore, we introduce the dimensionless parameter 
\begin{equation}\label{eq:alpha}
	\alpha = \frac{\gamma_0}{\gamma_{\infty}} -1\ge 0, 
\end{equation}
in such a way that, for either $\alpha = 0$ or $\tau_0 \rightarrow 0$, the memory kernel becomes $K(t) = 2\gamma \delta(t)$, with constant friction coefficient $\gamma = \gamma_0 = \gamma_{\infty}$. Consequently, in these cases Equation~(\ref{eq:GLEnonstatT}) reduces to 
\begin{equation}\label{eq:LEnonstatT}
    \gamma \frac{dx(t)}{dt} = -\kappa(t) x(t) + \zeta(t),
\end{equation}
where the thermal noise $\zeta(t)$ simply satisfies~\citep{brey1990}
\begin{eqnarray}\label{eq:whitenoise}
    \langle \zeta (t)\rangle & = & 0,\nonumber\\
    \langle \zeta(t) \zeta(s) \rangle & = & 2 k_B T(t) \gamma \delta(t-s),
\end{eqnarray}
Equation~(\ref{eq:LEnonstatT}) describes the motion of a Brownian particle coupled to a viscous heat bath with time dependent temperature $T(t)$ through the frictional force $-\gamma\frac{dx(t)}{dt}$ and the thermal stochastic force, subject to a restoring force $-\kappa(t) x(t)$. It should be noted that this situation was explicitly considered in many of the models of single-particle heat engines reported in the literature~\citep{Rana2014,Tu2014,Zakine2017,Saha2018,Saha2019,Holubec2020,Ekeh2020,Kumari2020,Szamel2020}. 

We point out that the rheological properties of viscoelastic fluids are generally dependent on their temperature, which under a thermodynamic cycle would also become time-dependent. The inclusion of such thermal effects in the minimal Langevin model (\ref{eq:GLEnonstatTexp}) is not trivial and even a phenomenological description through additional rheological parameters and time-scales would render it little useful for a clear interpretation of the memory effects of a frequency-dependent friction in the performance of the Brownian engine. Therefore, similar to the simplifications made in most single-particle models of heat engines working in purely viscous fluids, as a first approximation we assume that  $\eta_0$, $\eta_{\infty}$ and $\tau_0$ remain constant over time. The effect of the temperature dependence of these parameters is out of the scope of the present paper and will be the subject of further work.

\begin{figure}[h!]
\begin{center}
\includegraphics[width=0.975\textwidth]{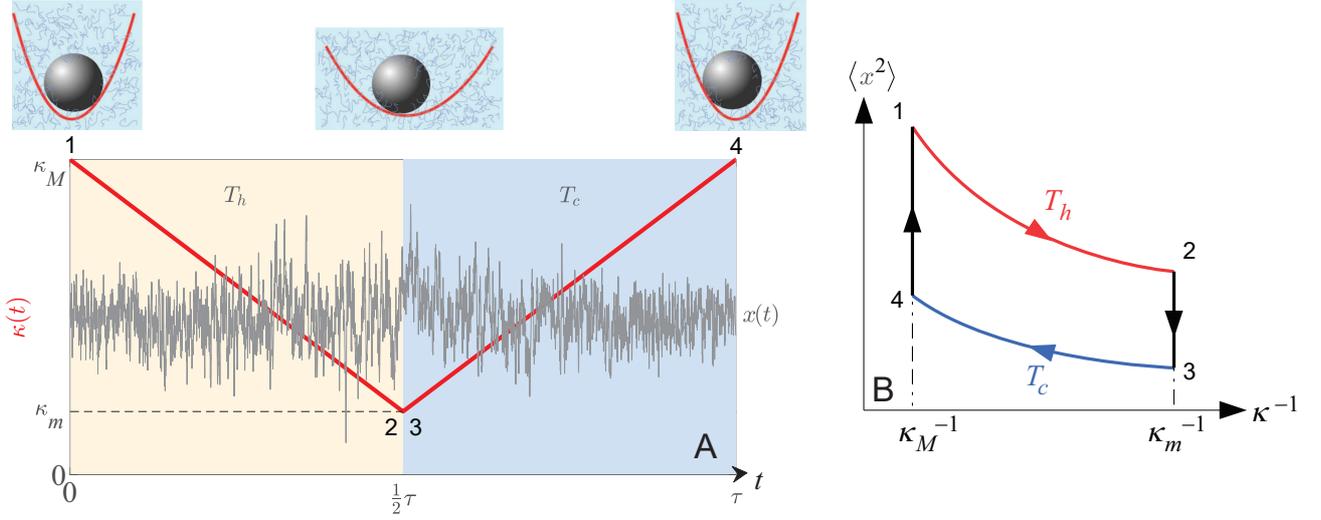}
\end{center}
\caption{{\bf{(A)}} Schematic representation of a Stirling cycle of period $\tau$ perfomed by a colloidal heat engine, embedded in a viscoelastic fluid, by means of the temporal variation of the stiffness $\kappa(t)$ of a trapping harmonic potential $U(x,t) = \frac{1}{2}\kappa(t) x(t)^2$ and of the bath  temperature $T(t)$. At time $0 \le t < \tau/2$, the trap stiffness is linearly decreased from $\kappa_M$ to $\kappa_m < \kappa_M$ while keeping the temperature of the surroundings at high temperature $T_h$ (step $1\rightarrow 2$). At $t = \tau/2$, the temperature is suddenly decreased to $T_c < T_h$ (step $2\rightarrow 3$), and kept at that value for $\tau/2 < t < \tau$, while linearly increasing the trap stiffness from $\kappa_m$ to $\kappa_M$ (step $3\rightarrow 4$). The cycle is completed at $t = \tau$, at which the temperature is again increased to $T_h$ (step $4 \rightarrow 1$). For a given realization of the cycle, the particle position $x(t)$ encodes the information of the stochastic energy exchange between the particle and the surrounding fluid, as depicted by the noisy trajectory obtained by numerical simulations of Equation~(\ref{eq:GLEnonstatTexp}). {\bf{(B)}} Schematic representation of the Brownian Stirling cycle in a $\langle x^2 \rangle$-$\kappa^{-1}$ diagram, similar to the pressure-volume diagram of a gas.}\label{Fig:1}
\end{figure}

The operation of the Brownian engine during a Stirling cycle of duration $\tau$ is depicted in Figure~\ref{Fig:1}(A), where the trap stiffness and the temperature are varied in time $t$ according to the following protocols
\begin{equation}\label{eq:Stirlingkappa}
    \kappa(t) = \left\{
    \begin{array}{ll}
    \kappa_M - \frac{2}{\tau} \delta \kappa   t, & \,\,\,\,\, 0 \le t \le \frac{\tau}{2},\\
    \kappa_m - \delta \kappa \left(1-\frac{2}{\tau}t \right), & \,\,\,\,\, \frac{\tau}{2} < t \le \tau,
    \end{array} \right.
\end{equation}
and
\begin{equation}\label{eq:StirlingT}
    T(t) = \left\{
    \begin{array}{ll}
    T_h, & \,\,\,\,\, 0 \le t < \frac{\tau}{2},\\
    T_c, & \,\,\,\,\, \frac{\tau}{2} \le t < \tau,\\
   T_h, & \,\,\,\,\,  t = \tau,
    \end{array} \right.
\end{equation}
respectively, where $\delta \kappa = \kappa_M - \kappa_m > 0$ and  $T_h > T_c$. More specifically, a full cycle consists of a sequence of four steps:
\begin{itemize}
	\item[$1\rightarrow 2$:] {For $0 \le t < \tau/2$, the colloidal engine undergoes an isothermal expansion at high themperature $T_h$ by linearly decreasing the trap stiffness from $\kappa_M$ to $\kappa_m$}.
	\item[$2\rightarrow 3$:]{At $t = \tau/2$, the temperature is suddenly decreased to $T_c$, while keeping the trap stiffness at $\kappa(t =\tau/2) = \kappa_m$, thus corresponding to a isochoric-like process.}
	\item[$3\rightarrow 4$:]{For $\tau/2 < t < \tau$, the engine undergoes an isothermal compression at low themperature $T_c$ by linearly increasing the trap stiffness from $\kappa_m$ to $\kappa_M$}.
	\item[$4\rightarrow 1$:]{At $t = \tau$, the temperature is suddenly raised to $T_h$, while keeping the trap stiffness at $\kappa(t = \tau) = \kappa_M$, i.e. an isochoric-like process, thus  completing the full cycle.}
\end{itemize}
Then, the cycle is repetead until the system reaches a time-periodic steady state, which becomes independent of the choice of the initial condition $x(t=0) = x_0$.  Note that, by analogy with a macroscopic Stirling cycle of a gas as a working substance, here the inverse of the trap stiffness and the variance of the particle position play the role of the volume and pressure, respectively, as depicted in Figure~\ref{Fig:1}(B).

According to stochastic thermodynamics~\cite{Seifert2012}, the work done on the system by the time variation of the optical trap over a single stochastic realization of the $(n+1)-$th cycle starting at $t_n = n\tau$, with $n = 0,1,2,..$, is
\begin{eqnarray}\label{eq:work}
	W_{\tau} & = & \frac{1}{2}\int_{t_n}^{t_n+\tau} dt \, \frac{d\kappa(t)}{dt} x(t)^2,\nonumber\\
	& = & \frac{\delta \kappa}{\tau} \left[ - \int_{t_n}^{t_n+\frac{\tau}{2}} dt \, x(t)^2 + \int_{t_n+\frac{\tau}{2}}^{t_n+\tau} dt \, x(t)^2  \right],
\end{eqnarray}
whereas the heat dissipated into the bath during the first half period of the cycle is given by
\begin{eqnarray}\label{eq:heat}
	Q_{\tau/2} & = & W_{\tau/2} - \Delta U_{\tau/2}\,\nonumber\\
	& = & - \frac{\delta \kappa}{\tau} \int_{t_n}^{t_n+\frac{\tau}{2}} dt \, x(t)^2 - \frac{1}{2}  \left[  \kappa_m x\left(t_n+\frac{\tau}{2}\right)^2 -  \kappa_M x(t_n)^2 \right].
\end{eqnarray}
In Equation~(\ref{eq:heat}), $W_{\tau/2}$ is the work done during the first half of the cycle, and $\Delta U_{\tau/2}$ is the corresponding variation of the potential energy in the harmonic trap, $U(x,t)$, in accordance with the stochastic extension of the first law of thermodynamics. 
Positive and negative values of $W_{\tau}$ correspond to work done on the particle and work performed by the particle, respectively, whereas positive and negative values of $Q_{\tau/2}$ represent heat transfered from the particle to the bath and heat absorbed by the particle, respectively.
It must be noted that the mean steady-state values of the two stochastic variables given by Equations (\ref{eq:work}) and (\ref{eq:heat}), which will be denoted as $\langle W_{\tau} \rangle $ and $\langle Q_{\tau/2} \rangle$, respectively, are the ones needed for the calculation of the efficiency of the Stirling heat engine~\citep{Schmiedl2007}. They involve the variance of the particle position at an arbitrary time $t \ge 0$, $\langle x(t)^2 \rangle$, with $t = 0$ the time defining the initial condition, computed over an ensemble of independent realizations of the colored nosie $\zeta(t)$ defined by Equations~(\ref{eq:colorednoise}).
An analytical treatment of this problem requires the explicit solution of the generalized Langevin equation~(\ref{eq:GLEnonstatTexp}), which is not trivial even in the simpler case of a constant trap stiffness and constant temperature~\citep{Di_Terlizzi2020}. Therefore, to address the problem of the performance of a Brownian Stirling heat engine described by Equations~(\ref{eq:colorednoise}), (\ref{eq:GLEnonstatTexp}), (\ref{eq:Stirlingkappa}) and (\ref{eq:StirlingT}), we opt for numerical simulations of the corresponding stochastic dynamics.

\subsection{Numerical solution}

In order to compute the probability distributions of the work and the heat defined in Equations (\ref{eq:work}) and (\ref{eq:heat}), as well as their corresponding mean values, the non-Markovian Langevin Equation (\ref{eq:GLEnonstatTexp}) must be numerically solved. To this end, we  express it in an equivalent Markovian form by introducing an auxiliary stochastic variable, $z(t)$, defined as
\begin{equation}\label{eq:z}
    z(t) = \frac{1}{\tau_0} \int_{0}^t ds\, \exp \left( - \frac{t-s}{\tau_0} \right)\sqrt{\frac{T(t)}{T(s)}}\left[x(s)+\tau_0 \sqrt{2\Delta(s)} \xi_z(s) \right],
\end{equation}
where
\begin{equation}\label{eq:Delta}
    \Delta(s) = \frac{k_B T(s)}{\gamma_{0} - \gamma_{\infty}},
\end{equation}
represents a diffusion coefficient associated to the effective friction $\gamma_0 - \gamma_{\infty}$, which depends on the instantaneous value of the temperature at time $s$, $T(s)$, and $\xi_z(s)$ is a Gaussian noise satisfying
\begin{eqnarray}\label{eq:noisez}
   \langle \xi_z (s)\rangle & = & 0,\nonumber\\
    \langle \xi_z(s) \xi_z(s') \rangle & = & \delta(s-s').
\end{eqnarray}
Consequently, the non-Markovian Langevin equation~(\ref{eq:GLEnonstatT}) for $x(t)$ can be written as a linear system of two coupled Markovian Langevin equations 
\begin{eqnarray}\label{eq:exponential2Markovian}
   \frac{dx(t)}{dt} & = & -\frac{\alpha + 1}{\gamma_0} \kappa(t) x(t) - \frac{\alpha}{\tau_0} \left[ x(t) - z(t) \right] + \sqrt{2 D_{\infty}(t)} \xi_x(t),\\
    \frac{dz(t)}{dt} & = & -\frac{1}{\tau_0} \left[ z(t) - x(t) \right] + \frac{1}{2T(t)}\frac{dT(t)}{dt}z(t) + \sqrt{2\Delta(t)} \xi_z(t),
\end{eqnarray}
with $\alpha$ defined in Equation (\ref{eq:alpha}). In Equation~(\ref{eq:exponential2Markovian}), $D_{\infty}(t)$ is a short-time diffusion coefficient associated to the infinite-frequency friction coefficient $\lim_{\omega \rightarrow \infty}\hat{K}^*(\omega) =\gamma_{\infty}$, see Equation~(\ref{eq:Fourierkernelexp}), at temperature $T(t)$, and is given by
\begin{equation}\label{eq:Dinfty}
    D_{\infty}(t) = \frac{k_B T(t)}{\gamma_{\infty}},
\end{equation}
whereas $\xi_x(t)$ is a Gaussian noise which satisfies
\begin{eqnarray}\label{eq:noisex}
   \langle \xi_x (t)\rangle & = & 0,\nonumber\\
    \langle \xi_x(t) \xi_x(s) \rangle & = & \delta(t-s),
\end{eqnarray}
Note that, apart from the step-like changes at $t = t_n$ and $t = t_n + \frac{\tau}{2}$, $T(t)$ remains constant. Accordingly, the rate of change of the time-dependent temperature in Equation (\ref{eq:exponential2Markovian}) vanishes during each half a Stirling cycle, i.e., $\frac{d}{dt}\left[ \ln T(t) \right]= 0$. 

To compute the probability distributions of $	W_{\tau}$ and $Q_{\tau/2}$, we carry out numerical simulations of the stochastic process $[x(t),z(t)]$ starting from the initial condition $[x(t=0) = 0,z(t=0) =0]$ with a total length of $2 \times 10^4$ times the period $\tau$. To ensure that the system is always in a time-periodic non-equilibrium steady state independent of the choice of the initial condition, the first $10^4$ cycles are left out and the origin of time is shifted to the beginning of the $(10^4+1)-$st cycle. Furthermore, without loss of generality  we choose constant values of the low and high-frequencies viscosities that are typical of viscoelastic fluids prepared in aqueous solution in semidilute regimes~\citep{Handzy2004,Zhu2008,Chapman2014,GomezSolano2015,Paul2021}: $\eta_0 = 0.040$~Pa~s and $\eta_{\infty} = 0.004$~Pa~s, which correspond to $\alpha = 9$. The diameter of the colloidal particle is set to $a = 0.5\,\mathrm{\mu}$m, while the maximum and minimum values of the trap stiffness during the Stirling cycle are chosen as $\kappa_M = 5\,\mathrm{pN}\,\mu\mathrm{m}^{-1}$ and $\kappa_m = 1\,\mathrm{pN}\,\mu\mathrm{m}^{-1}$, respectively, which are easily accessible with optical tweezers~\citep{gieseler2020}. The temperatures of the reservoir during the hot and cold part of the cycle are $T_c = 5^{\circ}$C and  $T_h =90^{\circ}$C,  which are selected in such a way that they are within the temperature range in which water, which is a common solvent component of many viscoelastic fluids, remains liquid. On the other hand, to study the influence of the fluid relaxation time on the performance of the colloidal Stirling engine, $\tau_0$ is varied in the range of 0.01~s$-$100~s, which also covers characteristic values in actual experimental systems. We solve Equations (\ref{eq:exponential2Markovian}) by means of an Euler–Cromer scheme with time step $\delta t = 10^{-4}$~s, which is about 75 times smaller than the shortest relaxation time of the system, $\gamma_{\infty}/\kappa_M$. 
In the case of the Stirling heat engine in a Newtonian viscous fluid, we solve numerically Equation (\ref{eq:LEnonstatT}) with constant friction coefficient $\gamma = 6\pi a \eta$, where $\eta = \eta_0 = 0.040$~Pa~s and the rest of the involved parameters, namely, $\kappa_m$, $\kappa_M$, $a$, $T_c$, $T_h$, and $\delta t$, are selected with the same values as described before for the viscoelastic case for a direct comparison between both systems. 
We also explore different values of the cycle period, 0.01~s~$\le \tau \le$~50~s, which allows us to examine the approach of the computed quantities to the quasi-static values $\tau \rightarrow \infty$. 
We note that $\tau_{\kappa} \equiv {\gamma_0/\kappa_m}$ represents the slowest dissipation time-scale of the system~\citep{Albay2021}, and appears explicitly in the analytical expressions for the variance of a Brownian particle undergoing a finite-time Stirling cycle in contact with a viscous heat bath~\citep{Kumari2020}. Therefore, in both cases of the viscous and viscoelastic baths analyzed here, all the timescales are normalized by $\tau_{\kappa}$, whereas energies are normalized by $k_B T_c$.

\section{Results and discussion}

\begin{figure}[h!]
\begin{center}
\includegraphics[width=1\textwidth]{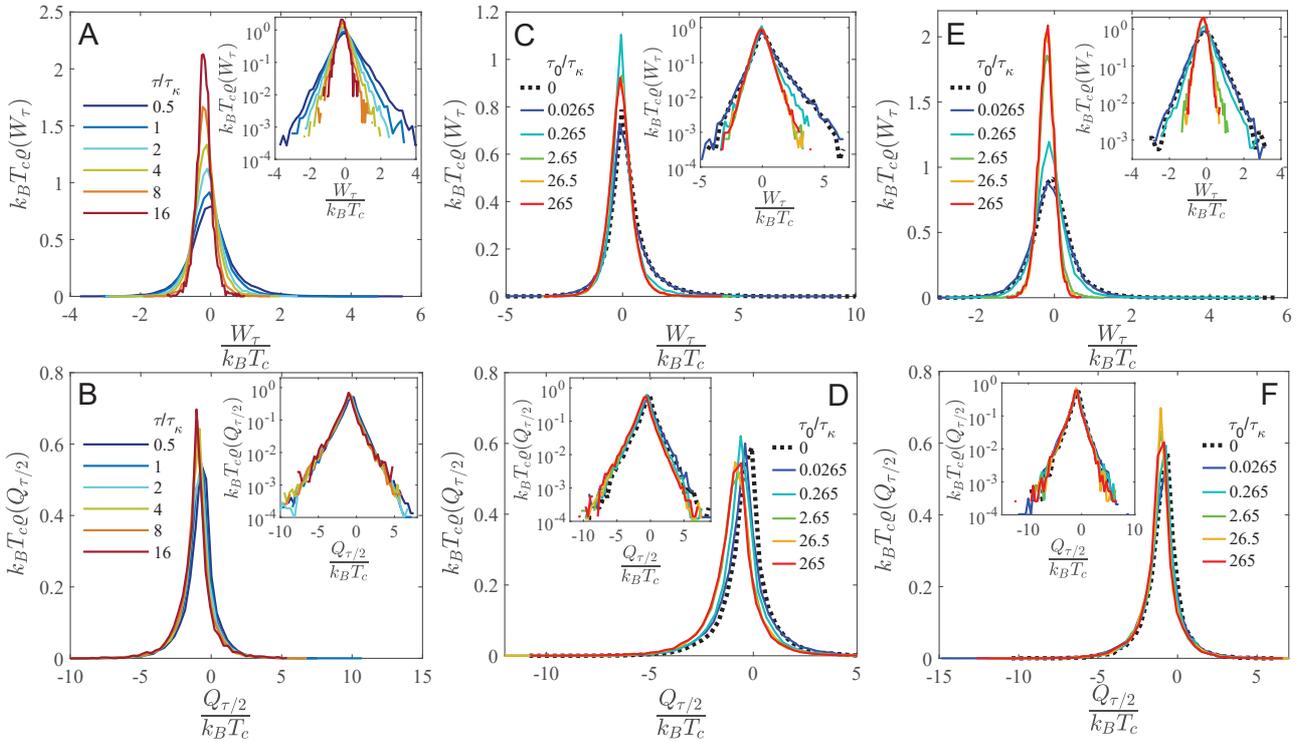}
\end{center}
\caption{ \textbf{(A)} Probability density function of the work $W_{\tau}$, and \textbf{(B)} the heat $Q_{\tau/2}$, for a Brownian Stirling engine in contact with a viscoelastic fluid bath with relaxation time $\tau_0 = 2.65 \tau_{\kappa}$, for different values of the cycle time $\tau$. \textbf{(C)} Probability density function of the work $W_{\tau}$, and \textbf{(D)} the heat $Q_{\tau/2}$, for a Brownian Stirling engine during a cycle of duration $\tau = \tau_{\kappa}$, in contact with viscoelastic fluid baths with the same zero-shear viscosity $\eta_0 = 0.040$~Pa~a, and distinct relaxation times $\tau_0$ spanning 5 orders of magnitude (solid lines). \textbf{(E)} Probability density function of the work $W_{\tau}$, and \textbf{(F)} the heat $Q_{\tau/2}$, for a Brownian Stirling engine during a cycle of duration $\tau = 10 \tau_{\kappa}$, in contact with viscoelastic fluid baths with the same zero-shear viscosity $\eta_0 = 0.040$~Pa~s and distinct relaxation times $\tau_0$ spanning 5 orders of magnitude (solid lines). In Figures \ref{Fig:2}\textbf{(C)}-\textbf{(F)}, the dotted lines represents the corresponding curves for a Brownian engine in a Newtonian fluid ($\tau_0 = 0$) with constant viscosity $\eta_0 = 0.040$~Pa~s. The insets are semilogarithmic representations of the main plots.}\label{Fig:2}
\end{figure}

Since $W_{\tau}$ and $Q_{\tau/2}$ are stochastic variables, we first present the results for their probability distributions, $\varrho(W_{\tau})$ and $\varrho(Q_{\tau/2})$, respectively, for different values of the time-scales $\tau$ and $\tau_0$. In Figure \ref{Fig:2}(A) and (B) we plot such distributions for a value of the fluid relaxation time that is comparable to the largest dissipation time-scale of the system:  $\tau_0 = 2.65 \tau_{\kappa}$, at which memory effects due to the frequency-dependent friction must be important. In such a case, we observe that for fast Stirling cycles with period  $\tau$ smaller or comparable to $\tau_{\kappa}$ the work distribution is asymmetric with respect to its maximum and exhibits pronounced exponential tails, as illustrated in the inset of Figure \ref{Fig:2}(A). In addition, large positive work fluctuations occur for small $\tau$, which indicates the existence of rare events where work is done on the particle during a cycle, thus effectively consuming energy as a heat pump. As $\tau$ increases, the exponential tails and their asymmetry vanish, thus giving rise to a narrower Gaussian-like shape for $\tau \gg \tau_k$. This shows that the probability of finding positive work fluctuations decreases by increasing $\tau$, i.e., the Brownian particle behaves more and more like a macroscopic Stirling engine, which on average is able to convert the heat absorbed from the viscoelastic bath into work. On the contrary, the heat distribution does not significantly change with the cycle time $\tau$, as shown in Figure \ref{Fig:2}(B). In this case, clear exponential tails remain even for large values of $\tau$, as revealed in the inset of Figure~\ref{Fig:2}(B). where the probability of occurence of negative heat fluctuations is higher than that of positive ones. Hence, regardless of the the cycle period $\tau$, it is more likely that heat is absorbed by the particle than dissipated into the bath during the isothermal expansion at temperature  $T_h$.

In Figures \ref{Fig:2}(C) and (D) we analyze the dependence on the fluid relaxation time $\tau_0$ of the work and heat distributions, respectively, for Stirling cycles of period $\tau = \tau_{\kappa}$, i.e., similar to the largest viscous dissipation time-scale of the system. For comparison, we also plot as dotted lines the corresponding probability distributions for a colloidal engine in a fluid with constant viscosity $\eta = \eta_0$, for which $\tau_0 = 0$. Remarkably, we find that the fluid viscoelasticity, through the parameter $\tau_0$, has a strong influence on the resulting shape of the distributions. For a viscous bath, the work has large exponential tails with a highly asymmetric shape. A similar shape is observed for a viscoelastic bath at sufficiently small $\tau_0$, but the width and the asymmetry of the distribution gradually decrease as $\tau_0$ increases, then converging to a single limiting curve with a rather symmetric profile for sufficiently large values of the fluid relaxation time $\tau_0 \gtrsim \tau_{\kappa}$, as shown in the inset of Figure \ref{Fig:2}(C). In addition, the heat distribution also has exponential tails with a width that does not strongly depend on the fluid relaxation time $\tau_0$, but the location of the maximum is slightly shifted to more and more negative values of $Q_{\tau/2}$ with increasing $\tau_0$, as shown in Figure \ref{Fig:2}(D). Finally, for values of the cycle duration $\tau$ larger than $\tau_{\kappa}$, the shape of $\varrho(W_{\tau})$ changes from a rather symmetric exponentially-tailed distribution to a limiting Gaussian curve with increasing $\tau_0$, whereas $\varrho(Q_{\tau/2})$ exhibits a symmetric profile with exponential tails peaked at a negative value of $Q_{\tau/2}$, which remains unaffected by the $\tau_0$, as respectively shown in Figures \ref{Fig:2}(E) and (F) for $\tau = 10 \tau_{\kappa}$. It is important to realize that for $\tau > \tau_{\kappa}$, the work distribution of the Brownian engine is narrower in a viscoelastic bath as compared to that in a viscous bath with the same zero-shear viscosity. This can be attributed the elastic response in the former case, which prevents large instantaneous heat losses into the bath by viscous dissipation, thus resulting in a more efficient conversion into work of the energy extracted from the surroundings. This observation underlines the importance of the friction memory kernel of the particle motion in the viscoelastic fluid, which becomes strongly dependent on the frequency imposed by the Stirling cycle. Thus, for sufficiently small $\tau_0 < \tau_{\kappa}$ the energy exchanges between the Brownian particle and the viscoelastic bath must  not be that different from those ocurring in a viscous fluid, while for sufficiently large $\tau_0 > \tau_{\kappa}$ significant deviations must take place, as verified in Figures \ref{Fig:2}(C) and (E) for $\tau_0 = 0.0265 \tau_{\kappa}$ and $\tau_0 = 265 \tau_{\kappa}$, respectively.

\begin{figure}[h!]
\begin{center}
\includegraphics[width=0.95\textwidth]{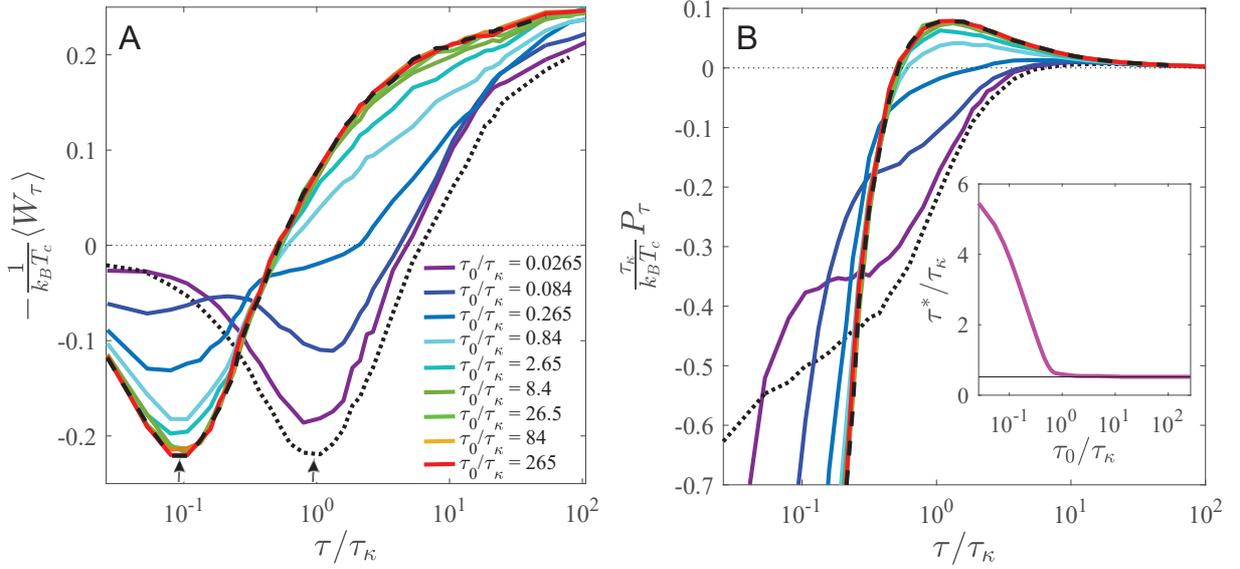}
\end{center}
\caption{{\textbf{(A)}} Mean work done by the Brownian Stirling engine during a cycle,  $-\langle W_{\tau} \rangle$, as a function of the cycle time $\tau$, for different values of the fluid relaxation time $\tau_0$ (solid lines). The dotted and dashed lines represent the mean power output of a Brownian Stirling engine in Newtonian fluids with constant viscosities $\eta = \eta_0$ and $\eta =(1+\alpha)^{-1} \eta_0 = \eta_{\infty}$, respectively. The arrows depict the location of the corresponding minima. {\textbf{(B)}} Mean power output per cycle of the colloidal Stirling engine, $P_{\tau}$, as a function of the cycle duration $\tau$, for different values of the fluid relaxation time $\tau_0$ (solid lines). The dotted and dashed lines represent the mean power output of a Brownian Stirling engine in Newtonian fluids with constant viscosities $\eta = \eta_0$ and $\eta =(1+\alpha)^{-1} \eta_0 = \eta_{\infty}$, respectively. Same color code as in Figure~\ref{Fig:3}(A). Inset: stall time of the engine defined in Equation~(\ref{eq:stalltime}), $\tau^*$, as a function of the fluid relaxation time, $\tau_0$ (thick line). The thin line represents the value of  $\tau^*$ for an engine operating in a Newtonian fluid with constant viscosity $\eta = \eta_{\infty}$.}\label{Fig:3}
\end{figure}

To investigate the performance of a Brownian engine operating in a viscoelastic bath, in Figure~\ref{Fig:3}(A) we plot the mean work \emph{done by the Brownian engine} during a cycle, i.e., $-\langle W_{\tau} \rangle$. In a Newtonian fluid,  $-\langle W_{\tau} \rangle$ is positive at sufficiently large $\tau$ and monotonically saturates to a constant positive value in the quasi-static limit $\tau \rightarrow \infty$~\citep{Kumari2020}
\begin{equation}\label{eq:qswork}
	-\langle W_{\tau\rightarrow \infty} \rangle = \frac{1}{2}k_B(T_h-T_c)\ln \left( \frac{\kappa_M}{\kappa_m} \right),
\end{equation}
whereas it becomes negative at small values of $\tau$ and tends to zero as $\tau \rightarrow 0$ according to Equation~(\ref{eq:work}), thus implying that it has a minimum at a certain value of $\tau$. This is verified in Figure~\ref{Fig:3}(A), where we plot as dotted and dashed lines the curves corresponding to the work done by a particle in viscous fluids with constant viscosities $\eta = \eta_0 = 0.040$~Pa~s and $\eta = \eta_{\infty} = 0.004$~Pa~s, respectively, i.e., equal to the viscosities characterizing the long-time and short-time dissipation of the viscoelastic fluid. The location of the minimum, which is depicted by arrows, depends on the specific value of $\eta$, but the general shape of the curve in a linear-logartihmic representation is the same, as observed in Figure~\ref{Fig:3}(A). Interestingly, in the case of viscoelastic fluids with non-zero values of $\tau_0$, the work done by the particle exhibits an intermediate behavior between these two curves. For instance, for $\tau_0 = 0.0265 \tau_{\kappa} \ll  \tau_{\kappa}$, the dependence of $-\langle W_{\tau} \rangle$ on $\tau$ is very similar to that in a Newtonian fluid with viscosity $\eta = \eta_0$, with a single minimum at the same location ($\tau \approx \tau_{\kappa}$) and only small deviations of the respective values along the vertical axis. Nevertheless, as $\tau_0$ increases, a second local minimum emerges at $\tau \approx 0.1 \tau_{\kappa}$, i.e., at the location of the minimum of the curve corresponding to the Newtonian fluid of viscosity $\eta = \eta_{\infty}$, as observed in Figure~\ref{Fig:3}(A) for $\tau_0 =  0.084 \tau_{\kappa}$. Such a second mininum becomes more and more apparent with increasing $\tau_0$, whereas the first minimum at $\tau \approx \tau_{\kappa}$ becomes less and less dominant, as seen for $\tau_0 \ge 0.265 \tau_{\kappa}$. Unexpectedly, for $\tau_0 \gg \tau_{\kappa}$, the curves for the viscoelastic case converge to that for a Newtonian fluid with a viscosity $\eta = \eta_{\infty}$. These observations suggest that, depending of the specific values of the fluid relaxation time and the cycle time with respect to $\tau_{\kappa}$, different dissipation mechanisms  take place in order for the particle to convert the energy taken from the bath into work by means of the applied thermodynamic cycle.

Next, we compute the mean power produced by the engine during a cycle
\begin{equation}\label{eq:power}
	P_{\tau} = -\frac{\langle W_{\tau} \rangle}{\tau},
\end{equation}
whose dependence on the cycle time $\tau$ is plotted as solid lines in Figure \ref{Fig:3}(B) for some exemplary values of the fluid relaxation time $\tau_0$. Besides, we also plot in Figure \ref{Fig:3}(B) as a dotted line the mean power for a Brownian engine in a Newtonian fluid bath with viscosity $\eta = \eta_0$. It is important to note that, for all values of $\tau_0$, $P_{\tau}$ exhibits a non-monotonic behavior as a function of $\tau$, which gradually deviates from the behavior in a Newtonian fluid with viscosity $\eta = \eta_0$ as $\tau_0$ increases. This is the result of the pronounced non-monotonic dependence of $-\langle W_{\tau} \rangle$ on $\tau$ shown in Figure~\ref{Fig:3}(A). In particular, $P_{\tau}$ has a maximum that originates from the trade-off between high energy dissipation at small cycle times $\tau$ (high frequency operation) and large $\tau$ (slow operation), at which net work is produced by the engine with low dissipation. Additionally, the general shape of all power curves displays three different operation regimes. For sufficiently slow Stirling cycles (large $\tau$), the engine is able to deliver net power on average ($P_{\tau} >0$), where the irreversible energy dissipation into the bath becomes negligible. On the other hand, there is a specific value of the cycle time at which the engine stalls, i.e., both the mean work and the power output vanish: $\langle W_{\tau}\rangle = 0$, $P_{\tau} = 0$ \citep{Schmiedl2007}. Finally, for sufficiently fast cycles (small $\tau$), the engine absorbs energy ($P_{\tau} < 0$) rather than delivering it, thus behaving like a heat pump. This regime is the consequence of the large amount of energy irreversibly dissipated when the particle is quickly driven by the periodic variation of $\kappa(t)$ and $T(t)$. Interestingly, in Figure \ref{Fig:3}(B), we show that the value of the fluid relaxation time $\tau_0$ has a considerable impact on the mean power output, and in particular, on the value of the cycle time at which the Brownian engine stalls, which we denote as $\tau^*$
\begin{equation}\label{eq:stalltime}
	P_{\tau^*} = 0.
\end{equation}
For instance, in the case of the Newtonian fluid ($\tau_0=0)$ with $\eta = \eta_0$, we find $\tau^* = 6.15\tau_{\kappa}$, while for a viscoelastic fluid ($\tau_0 > 0$), $\tau^*$ is smaller and decreases with increasing $\tau_0$. In the inset of Figure \ref{Fig:3}(B) we plot the dependence of $\tau^*$ on $\tau_0$,  where we can see that for sufficiently short fluid relaxation times, the stall time is close to that for a Newtonian fluid bath ($\tau^* = 6.15\tau_{\kappa}$), and monotonically decreases with increasing $\tau_0$. In this short-$\tau_0$ regime, the performance of the engine is very sensitive to the specific value of $\tau_0$, as shown by the strong variation of the shape of the power curves plotted in Figure \ref{Fig:3}(B) for $\tau_0 = 0.0265\tau_{\kappa}, 0.084\tau_{\kappa},  0.265\tau_{\kappa}, 0.84\tau_{\kappa} $. Around $\tau_0 = \tau_{\kappa}$, a conspicuous change in the dependence on $\tau_0$ of the operation of the engine happens. Indeed, as $\tau_0$ increases the stall time converges to the constant value $\tau^* = 0.52 \tau_{\kappa}$, as verified in the inset of Figure \ref{Fig:3}(A) for $\tau_0 >  \tau_{\kappa}$. The monotonic decrease of $\tau^*$ implies that the interval of cycle times at which the engine is able to efficiently deliver positive power output is expanded with increasingly larger $\tau_0$. Moreover, with increasing fluid relaxation times $\tau_0 > \tau_{\kappa}$, which is consistent with increasingly pronounced viscoelastic behavior of the bath, the power output curves converge to a limiting curve, as shown in Figure \ref{Fig:3}(B)  for $\tau_0 = 2.65\tau_{\kappa}, 8.4\tau_{\kappa},   26.5\tau_{\kappa},  84\tau_{\kappa}, 265 \tau_{\kappa}$. Remarkably, we find that such a limiting curve corresponds to the power curve of a Brownian Stirling engine in a Newtonian bath with viscosity equal to high frequency value $\eta = \eta_{\infty} = 0.004$~Pa~s, i.e., the viscosity of the solvent component in the viscoelastic fluid, which is represented as a dashed line in Figure~\ref{Fig:3}(B). As a consequence, the limit of the stall time of an engine working in a viscoelastic fluid with increasing $\tau_0$ corresponds to the stall time of a Brownian engine operating in a Newtonian one with constant viscosity $\eta = \eta_{\infty}$, $\tau^* = 0.52 \tau_{\kappa}$, as verified in the inset of Figure \ref{Fig:3}(B), see the horizontal solid line. Furthermore, in Figure \ref{Fig:3}(B) we check that, for a given cycle of finite duration $\tau$, the mean power output of the engine operating in a viscoelastic fluid is enhanced with increasing values of $\tau_0$ with respect to the power output in a Newtonian fluid of the same zero-shear viscosity. We also find that the location of the global maximum of each power output is shifted to smaller and smaller values of $\tau$ with increasing $\tau_0$, whereas the value of $P_{\tau}$ at the maximum increases with increasing $\tau_0$ because of the decreasing irreversible dissipation taking place in a fluid with pronounced viscoelastic behavior.

These findings allows us to uncover the underlying mechanism behind the influence of fluid viscoelasticity on the performance of the engine. In a Newtonian fluid with constant viscosity $\eta_0$, the largest time-scale associated to viscous dissipation due to temporal changes in the trap stiffness is precisely $\tau_{\kappa}$, which is proportional to $\eta_0$, and represents the largest relaxation time in the system. In this case, the viscous bath simply acts as a mechanically inert element of the engine which equilibrates instaneously in response to the particle motion under the variations of the trap stifness. On the other hand, when the bath is a viscoelastic fluid, the hidden degrees of freedom of its elastic microstructure, e.g., entangled micelles, polymers, interacting colloids, etc., also come into play in the dynamics and mechanically respond within a characteristic time $\tau_0 > 0$ to the temporal changes periodically imposed on the particle. Therefore, the interplay between $\tau_{\kappa}$ and $\tau_0$ determines the resulting energetic behavior of the system: 
\begin{itemize}
	\item If $\tau_0 \ll \tau_{\kappa}$, the fluid microstructure fully relaxes before the energy dissipation into the bath takes place on a time-scale $\tau_{\kappa}$. In such circumstances, the Brownian particle has enough time to probe the long-time (low frequency) properties of the fluid environment with friction coefficient $ \hat{K}^*(\omega \rightarrow 0) = \gamma_0 = 6\pi a \eta_0$, see Equation (\ref{eq:Fourierkernelexp}), thereby leading to a stochastic energetic behavior similar to that in a Newtonian fluid with constant viscosity $\eta = \eta_0$. 
	\item If $\tau_0 \lesssim \tau_{\kappa}$, excessive irreversible energy losses by viscous dissipation are counterbalanced by the transient energy storage in the elastic structure of the bath, because at frequencies $\omega \sim \tau_0^{-1}$ the imaginary part of $\hat{K}^*(\omega)$ is not negligible. Therefore, the value $\tau_0 \approx \tau_{\kappa}$ marks a qualitative change in the energy exchange between the particle and bath. 
	\item If $\tau_0 > \tau_{\kappa}$, the elastic fluid microstructure does not have enough time to mechanically relax to the temporal changes of the cycle, thus preventing the particle from undergoing the long-time friction characterized by the coefficient $\gamma_0$. Therefore, the particle can only probe the short-time response of the surrounding fluid through the high-frequency components of the friction, which correspond to $\hat{K}^*(\omega \rightarrow \infty) = \gamma_{\infty} = 6\pi a \eta_{\infty}$ for $\tau_0 \gg \tau_{\kappa}$ according to Equation (\ref{eq:Fourierkernelexp}). As a consequence, in this limit the relevant dissipation timescale is $\gamma_{\infty}/\kappa_m$, which is in general smaller than $\tau_{\kappa}$ because $\eta_{\infty} = \eta_0(1+\alpha)^{-1} \le \eta_0$. For instance, for the numerical values chosen in the simulations presented here, $\gamma_{\infty}/\kappa_m = 0.1\tau_{\kappa}$. Accordingly, less irreversible dissipation must take place in the viscoelastic fluid under finite-time Stirling cycles, thus enhancing the net power output of the engine at a given cycle time $\tau$ as compared to that in a Newtonian fluid with the same zero-shear viscosity $\eta_0$.  
\end{itemize}

\begin{figure}[h!]
\begin{center}
\includegraphics[width=0.95\textwidth]{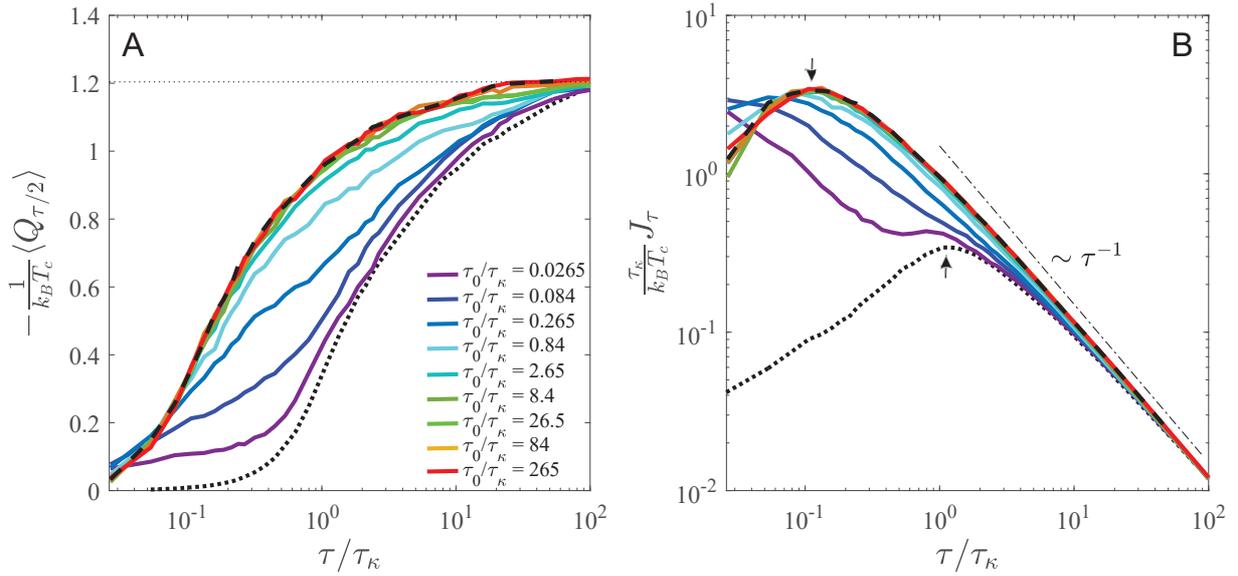}
\end{center}
\caption{{\textbf{(A)}} Mean heat absorbed by the Brownian Stirling engine during the isothermal expansion at high temperature,  $-\langle Q_{\tau/2} \rangle$, as a function of the cycle time $\tau$, for different values of the fluid relaxation time $\tau_0$ (solid lines). The dotted and dashed lines corresponds to the mean heat absorbed by the colloidal engine in Newtonian fluids with constant viscosities $\eta = \eta_0$ and $\eta =(1+\alpha)^{-1} \eta_0 = \eta_{\infty}$, respectively. The horizontal thin dotted line represents the quasi-static value given by Equation \ref{eq:qsheat}. \textbf{(B)}  Mean rate of heat absorption by the colloidal engine during the isothermal expansion at high temperature, $J_{\tau}$, as a function of the duration of  the cycle $\tau$, for different values of the fluid relaxation time $\tau_0$ (solid lines). Same color code as in \ref{Fig:4}(A). The dotted and dashed lines correspond to mean rate of heat absorption in Newtonian fluids with constant viscosities $\eta = \eta_0$ and $\eta = \eta_{\infty}$, respectively. The arrows depict the location of the corresponding maxima. The dotted-dashed line depicts the behavior $\sim \tau^{-1}$.}\label{Fig:4}
\end{figure}

To confirm the previously described mechanism of energy storage and dissipation during the Stirling cycle, in Figure~\ref{Fig:4}(A), we plot the mean heat absorbed by the particle during the hot step of the cycle, $-\langle Q_{\tau/2}\rangle$, as a function of the total duration $\tau$ of a full cycle. We find that, for all values of $\tau$ and of the fluid relaxation time $\tau_0$, $-\langle Q_{\tau/2}\rangle \ge 0$, which means that the particle absorbs heat on average during the first half of the cycle. In particular, for a given $\tau_0$ the mean absorbed heat increases monotonically from the value $-\langle Q_{\tau=0}\rangle = 0$, and saturates to a constant value corresponding to a quasi-static process as $\tau \rightarrow \infty$. For comparison, in Figure \ref{Fig:4}(A) we also plot as a dotted line the mean heat absorbed by the Brownian engine when operating in a Newtonian fluid with viscosity $\eta = \eta_0$. In such a case, it can be readily demonstrated from Equation (\ref{eq:heat}) that  $-\langle Q_{\tau/2}\rangle$ actually approaches a quasi-static value, which is explicitly given by~\cite{Kumari2020}
\begin{equation}\label{eq:qsheat}
	-\langle Q_{(\tau \rightarrow \infty)/2}\rangle = \frac{1}{2}k_B (T_h - T_c) + \frac{1}{2}k_B T_h \ln \left( \frac{\kappa_M}{\kappa_m}, \right)
\end{equation}
For the numerical values of the parameters investigated here, $-\langle Q_{(\tau \rightarrow \infty)/2}\rangle = 1.203 k_B T_c$, see horizontal thin dotted line in Figure \ref{Fig:4}(A). We observe that, regardless of $\tau_0$, all heat curves converge to such a value for $\tau \gg \tau_{\kappa}$, but depending on the specific value of the fluid relaxation time, different behaviors occur at short and intermediate cycle durations. Once again, we find that with increasing $\tau_0$, the  mean-heat curves gradually deviate from the behavior in a Newtonian fluid with viscosity $\eta  = \eta_0$, and for $\tau_0 \gg \tau_{\kappa}$ they converge to that in a Newtonian fluid with $\eta = \eta_{\infty}$, see dashed line in Figure~\ref{Fig:4}(A). This provides another evidence that, as $\tau_0$ increases, the energy dissipation of an engine operating in a viscoelastic fluid is mainly determined by the friction with the solvent.

In Figure~\ref{Fig:4}(B) we plot as solids lines the mean rate of heat absorption by the engine from the bath during the isothermal expansion at temperature $T_h$
\begin{equation}\label{eq:heatrate}
	J_{\tau} = -\frac{\langle Q_{\tau/2} \rangle}{\tau},
\end{equation}
as a function of the cycle duration $\tau$ for some representative values of $\tau_0 > 0$. The corresponding curves for a particle in Newtonian fluids with $\eta = \eta_0$ and $\eta = \eta_{\infty}$ are represented as dotted and dashed lines, respectively. In such cases, we find that $J_{\tau}$ exhibits a maximum, which corresponds approximately to the location of the minima in $-\langle W_{\tau} \rangle$ shown in Figure~\ref{Fig:3}(A). For $\tau \lesssim \tau_{\kappa}$, a marked dependence on the fluid relaxation time is observed if $\tau_0 \lesssim \tau_{\kappa}$, while for $\tau \gg \tau_{\kappa}$ a dependence $J_{\tau} \sim \tau^{-1}$ on the Stirling cycle time emerges for all values of $\tau_0$, thus indicating the onset of the quasi-static thermodynamic behavior.

The previous findings reveal that, unlike the performance of Brownian heat engines in a Newtonian environment with a single relevant time-scale $\gamma_0/\kappa_m$ of energy dissipation, in a viscoelastic fluid bath the low-frequency and the high-frequency values of the friction, $\gamma_0$ and $\gamma_{\infty}$, give rise two meaningful dissipation time-scales, namely $\tau_{\kappa} = \gamma_0/\kappa_m$ and the apparently hidden time-scale $(1+\alpha)^{-1}\tau_{\kappa} = \gamma_{\infty}/\kappa_m$ due to the friction of the particle with the solvent. When the Stirling cycle time $\tau$ is comparable to one of such time-scales, the corresponding channel of irreversible dissipation is strongly activated. This in turn leads to a large amount of energy absorbed by the particle from the heat bath at a very high rate, as manifested by the minima and maxima depicted by arrows in Figures~\ref{Fig:3}(A) and \ref{Fig:4}(B), respectively. For an arbitrary cycle time, the interplay between the two channels of irreversible dissipation along with the transient energy storage by the elastic microstructure of the fluid determine the resulting perfomance of the Brownian engine.

\begin{figure}[h!]
\begin{center}
\includegraphics[width=0.975\textwidth]{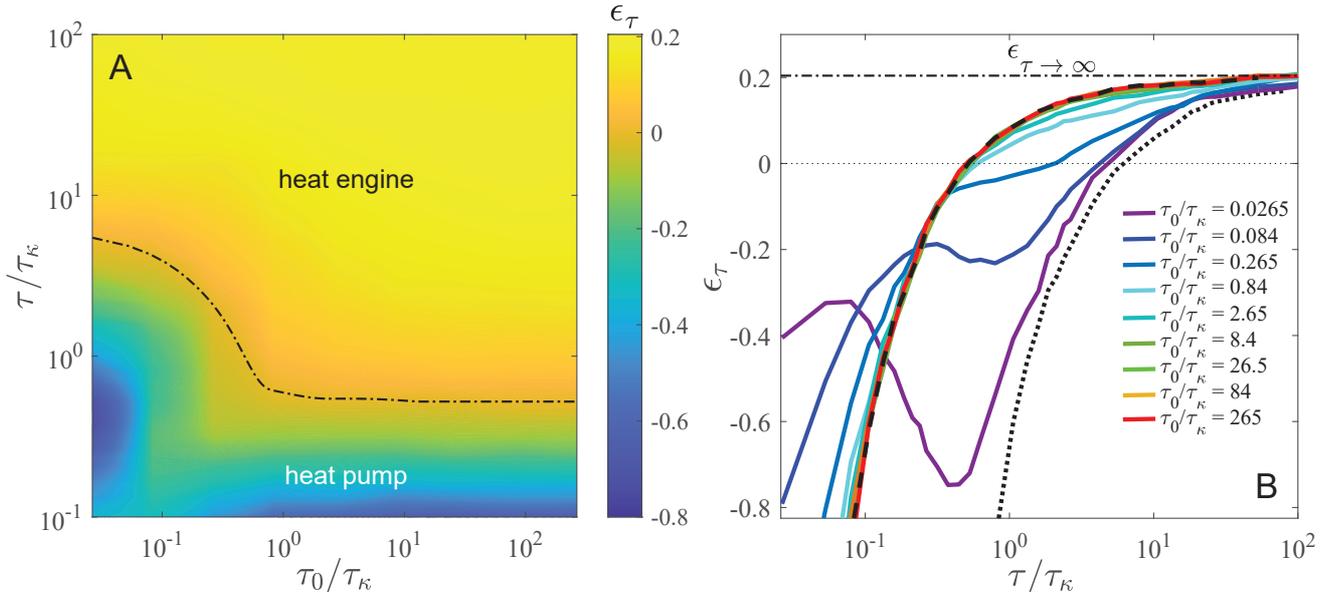}
\end{center}
\caption{\textbf{(A)} 2D color map representation of the efficiency of the colloidal Stirling engine, $\epsilon_{\tau}$, as a function of the fluid relaxation time, $\tau_0$, and the duration of a Stirling cycle, $\tau$. The dotted-dashed line corresponds to the stall time $\tau^*$, which separates the values of the parameters  $\tau_0$ and $\tau$ for which the system operates as a heat engine ($\epsilon_{\tau} >0$) from those for which it behaves as a heat pump ($\epsilon_{\tau} < 0$). \textbf{(B)} Examples of efficiency curves as function of the duration of a Stirling cycle, $\tau$, for some particular values of the fluid relaxation time (solid lines). The dotted and dashed lines represent the efficiency curves of a Brownian Stirling engine in Newtonian fluids with constant viscosity $\eta = \eta_0$ and $\eta = \eta_{\infty}$, respectively. The dotted-dashed line depicts the quasi-static value of the Stirling efficiency, $\epsilon_{\tau \rightarrow \infty}$, given by Equation \ref{eq:qsefficiency}.}\label{Fig:5}
\end{figure}

Finally, we determine the efficiency of the Brownian Stirling engine, defined as  
\begin{equation}\label{eq:Stirlingeff}
	\epsilon_{\tau} = \frac{-\langle W_{\tau}\rangle}{-\langle Q_{\tau/2} \rangle},
\end{equation}
as a function of the cycle time, $\tau$, and the relaxation time of the viscoelastic fluid, $\tau_0$. The results are represented as a 2D color map in Figure \ref{Fig:5}(A), with some efficiency curves plotted in Figure \ref{Fig:5}(B) as a function of $\tau$ for exemplary values of $\tau_0$. Additionally, in Figure \ref{Fig:5}(A) we also plot the stall time $\tau^*$ defined in Equation~(\ref{eq:stalltime}) as a function of the fluid relaxation time. As a consequence of the energy exchange  with a viscoelastic bath discussed in the previos paragraphs, $\tau^*$ divides the efficiency diagram into two regions. For $\tau < \tau^*$ the Brownian particle behaves a heat pump, where  $\epsilon_{\tau} < 0$ exhibits a rather intricate dependence on $\tau_0$ and $\tau$ due to the competition between the different energy storage and dissipation channels of the bath, which results on average in net energy absorption from the bath. On the other hand, for $\tau > \tau^*$, the efficiency is positive, $\epsilon_{\tau} >  0$, i.e., the Brownian particle behaves as a heat engine with positive power output. In this case, the efficiency is a monotonic increasing function of both $\tau$ and $\tau_0$. Note that for the investigated values of the cycle time $\tau$, the interval at which the engine has a positive efficiency is rather narrow for small fluid relaxation times $\tau_0 < \tau_{\kappa}$, because for the large value of the zero-shear viscosity considered in the simulations ($\eta_0 = 0.040$~Pa~s, typical of biological fluids), there is a large amount of heat dissipation even at comparatively slow Stirling cycles. However, when the value of $\tau_0$ is similar or larger than $\tau_{\kappa}$, the elastic response of the fluid takes effect, hence the decrease in energy dissipation with a subsequent broadening of the interval of cycle times by one order magnitude for which $\epsilon_{\tau} > 0$.

Because in the model (\ref{eq:GLEnonstatT}) we assume that the only source of stochasticity of the system is the thermal fluctuations of the fluid, apart from the driving potential of the harmonic trap there are no other sources of energy that affect the performance of the Brownian engine. Therefore, it is expected that the quasi-static Stirling efficiency 
\begin{equation}\label{eq:qsefficiency}
	\epsilon_{\tau \rightarrow \infty} = \frac{\epsilon_C}{1+\frac{\epsilon_C}{\ln\left(\frac{\kappa_M}{\kappa_m}\right)}},
\end{equation}
which can be determined from the ratio of Equations (\ref{eq:qswork}) and (\ref{eq:qsheat}), is never exceeded at finite $\tau$ regardless of the relaxation time of the viscoelastic fluid. In Equation (\ref{eq:qsefficiency}), $\epsilon_ C = 1 - \frac{T_c}{T_h}$ corresponds to the efficiency of a Carnot engine operating quasi-statically between two reservoirs at temperatures $T_c$ and $T_h$. For the numerical values of the parameters characterizing the Stirling cycle considered here, we find $\epsilon_{\tau \rightarrow \infty} = 0.2043$. In Figure \ref{Fig:5}(B) we demonstrate that, indeed, all the efficiency curves are bounded by such a value and approach it as the cycle time $\tau$ increases. The typical value of cycle period at which such efficiency is reached strongly depends on $\tau_0$. While for a Brownian engine in a Newtonian fluid of viscosity $\eta_0$ the convergence is very slow, the quasi-static Stirling efficiency can be reached in a viscoelastic bath for typical experimental values of the parameters of the system, as shown in Figure \ref{Fig:5}(B) for $\tau_0 > \tau_{\kappa}$.

To compare the performance of a Stirling Brownian engine in a viscoelastic bath with other situations of practical interest, we first determine its efficiency at maximum power in a Newtonian fluid bath with the same zero-shear viscosity $\eta = \eta_0$.  Although not as general as the Carnot efficiency, under some circumstances the so-called Curzon-Ahlborn efficiency~\citep{Novikov1958,Curzon1975} represents a good approximation for the upper bound of the efficiency of stochastic heat engines working at maximum power~\citep{VanDenBroeck2005,Schmiedl2007,Esposito2009,Holubec2014}. For the values of the parameters investigated in this work, we find that the power output $P_{\tau}$ in a purely viscous fluid reaches the maximum value $P_{\tau_{MP}} = 0.00679 {k_B T_c}{\tau_{\kappa}}^{-1}$ at a cycle time of $\tau_{MP} = 18. 6 \tau_{\kappa}$, at which the efficiency is $\eta_{\tau_{MP}} = 0.1218$. This value compares well with the Curzon-Ahlborn efficiency, $\epsilon_{CA} = 1 - \sqrt{\frac{T_c}{T_h}} =0.1248$, and is approximately 60\% the Carnot efficiency $\eta_C = 0.2043$. In Table~\ref{tab:efficiency} we list some exemplary values of the mean power output over a cycle, $P_{\tau = \tau_{MP}}$ , and the corresponding efficiencies of the Brownian engine, $\epsilon_{\tau = \tau_{MP}}$, operating at the same Stirling cycle time $\tau_{MP} = 18. 6 \tau_{\kappa}$ in viscoelastic fluid baths with distinct values of their relaxation time $\tau_0$. We verify that with increasing $\tau_0$, both the absolute power delivered by engine and its efficiency are enhanced with respect to those in a Newtonian fluid. In particular, the efficiency at $\tau_{MP} = 18. 6 \tau_{\kappa}$ converges to approximately 93\% the Carnot efficiency for $\tau_0 \gg \tau_{\kappa}$.

\begin{table}[h]
\caption{\label{tab:efficiency}Mean power output produced by a Brownian Stirling engine during a cycle $\tau = \tau_{MP} = 18. 6 \tau_{\kappa}$ and corresponding efficiency for distinct values of the fluid relaxation time $\tau_0$. In all cases, the zero shear viscosity is the same, $\eta_0 = 0.040$~Pa~s.}
\centering
{\renewcommand{\arraystretch}{1.5}\renewcommand{\tabcolsep}{0.2cm}
\begin{tabular}{|c|c|c|}

\hline
 $\tau_0 / \tau_{\kappa}$ &$\frac{\tau_{\kappa}}{k_B T_c}P_{\tau = \tau_{MP}}$ &$\epsilon_{\tau = \tau_{MP}}$ \\
\hline

0 & $ 0.00679$ & $0.1218$ \\ 
\hline
0.1 & $0.00867$ & $0.1484$  \\ 
\hline
1 & $0.00994$ & $0.1667$  \\ 
\hline

10 &$0.01142$ & $0.1829$  \\
\hline
100 &$0.01183$ & $0.1893$  \\ 
 \hline
\end{tabular}}
\end{table}

\section{Summary and final remarks}

In this work, we have investigated a stochastic model based on the generalized Langevin equation for a Brownian Stirling engine in contact with a viscoelastic fluid bath. The slow rheological behavior of the fluid is taken into account in the model by an exponentially decaying memory kernel, which captures the basic features of the linear viscoelastic behavior of many non-Newtonian fluids. Our findings demonstrate that the memory friction exerted by the surrounding fluid has a tremendous impact on the performance of the heat engine in comparison with its operation in a viscous environement with the same zero-shear viscosity. In particular, a pronounced enhancement of the power output and the efficiency of the engine occurs as a result of the frequency-dependent response of the fluid under finite-time Stirling cycles, thus converging to limiting curves determined by the high frequency component of the friction of the particle as the fluid relaxation time increases. Moreover, the minimum value  of the duration of the Stirling cycle at which the Brownian engine can convert energy from the medium into work becomes monotonically shorter with increasing fluid relaxation time, which broadens the interval of possible values of the Stirling cycle duration over which the engine is able to efficiently deliver positive power. From a wider perspective, our results highlight the importance of the non-equilibrium transient nature of the particle friction under temporal cycles of finite duration. We point out that, although in a different context, qualitatively similar effects have been discussed in systems with frequency-dependent properties due to their coupling to non-Markovian baths, such as Brownian particles driven into periodic non-equilibrium steady states~\citep{Wulfert2017} and quantum Otto refrigerators~\citep{Camati2020}. Furthermore, the link between a frequency dependent friction and the noise correlations of the bath is in turn an important issue for the correct interpretation of the efficiency of stochastic heat engines operating in nonequilibrium baths, as recently examined in the case of underdamped active Brownian particles~\citep{Holubec_2020}.

To the best of our knowledge, our work represents the first investigation on the effect of memory friction in the perfomance of a Brownian Stirling engine in contact with a viscoelastic fluid reservoir. Thus, we expect that the results presented in this paper will contribute to a better understanding and potential applications of efficient work extraction and heat dissipation in other types of mesoscopic engines operating in complex fluids. Further steps of our work aim at  addressing long-term memory effects during stochastic thermodyamic cycles with finite period, as those described by streched exponentials~\citep{Cui2017} and power law kernels and fractional Brownian noise~\citep{Qian2003,Rodriguez2015,Sevilla2019,GomezSolano2020}, which describe the mechanical response of diverse soft matter systems such as glasses and biological materials~\citep{Balland2006,Kobayashi2017}. One further aspect that could be investigated in the future is the effect of temporal changes in the fluid parameters, as it is well known that the rheological properties of viscoelastic fluids are dependent on their temperature, which under a thermodynamic cycle would become time-dependent. We would like to point out that, since the parameters characterizing the operation of the heat engine presented in this paper are representative of typical soft matter systems, we expect that this process can be realized in a straightforward manner by use of optical tweezers~\citep{gieseler2020}. Similar ideas could be extended to Brownian particles in non-linear potentials~\citep{Ferrer2021}, and active Brownian heat engines~\citep{Holubec2020} functioning in complex fluids, which could be implemented in practice by. e.g. light-activated colloids in non-Newtonian liquids~\citep{GomezSolano2017,GomezSolano2020SM,Narinder2018,Narinder2019,Lozano2019} and hot Brownian particles~\citep{Rings2010,Rings2012,Kumar2020}.





\section*{Conflict of Interest Statement}

The author declares that the research was conducted in the absence of any commercial or financial relationships that could be construed as a potential conflict of interest.

\section*{Author Contributions}
J.R.G.-S., concieved the model, carried out the numerical simulations, analyzed the results, and wrote the manuscript.

\section*{Funding}
This work was supported by UNAM-PAPIIT IA103320.

\section*{Data Availability Statement}
The datasets generated for this study are available on request to the corresponding author.


\end{document}